\documentclass[12pt]{article}
\usepackage[dvips]{color}
\usepackage{epsfig}
\usepackage{amsmath}
\usepackage{graphicx}

\begin{document}
\title{\bf Quantum  Fluctuations of  a   BTZ Black Hole in  Massive Gravity}
\author{{Behnam Pourhassan$^{a}$\thanks{Email: b.pourhassan@du.ac.ir}\hspace{1mm}, Mir Faizal$^{b,c}$\thanks{Email: mirfaizalmir@googlemail.com}\hspace{1mm}, Zaid Zaz$^{d}$\thanks{Email: mohammadzaz@gmail.com}\hspace{1mm}, and  Anha Bhat$^{e}$\thanks{Email:
anhajan1@gmail.com}}\\
$^{a}${\small {\em  School of Physics, Damghan University, Damghan, 3671641167, Iran}}\\
$^{b}$ {\small {\em  Irving K. Barber School of Arts and Sciences, University of British Columbia,}}\\
{\small {\em  Kelowna, British Columbia, V1V 1V7, Canada.}}\\
$^{c}$ {\small {\em  Department of Physics and Astronomy, University of Lethbridge,}}\\
{\small {\em  Lethbridge,   Alberta, T1K 3M4, Canada.}}\\
$^{d}$ {\small {\em  Department of Electronics and Communication Engineering, Institute of Technology,}}\\
{\small {\em  University of Kashmir, Srinagar, Kashmir-190006, India.}}\\
$^{e}$ {\small {\em  Department of Metallurgical and Materials Engineering, National Institute of Technology,}}\\
{\small {\em  Srinagar, Kashmir-190006, India.}}} \maketitle
\begin{abstract}
In this work, we shall analyze the effects of quantum fluctuations on the properties of a   BTZ black hole, in a massive theory of gravity.
We will analyze this for a charged BTZ black hole in   asymptotically AdS and dS space-times. The quantum fluctuations would produce thermal
fluctuations in the thermodynamics of this BTZ black hole. As these fluctuations would become relevant   at a sufficiently small scale,
we shall discuss the effects of such thermal fluctuations on the entropy of a    small charged BTZ black.  We shall also analyze the effects of
these fluctuations on the stability of such a black hole.
\end{abstract}
The second law of thermodynamics seems to be in conflict with the physics of black holes, if a maximum entropy is not associated with black holes \cite{1,2}.  This is because a spontaneous reduction in the entropy of the universe would occur by an object with finite entropy crossing the event horizon of a black hole.
So, a condition   for the second law of thermodynamics and black holes to co-exist is the association of maximum entropy with black holes. This   entropy associated with black holes is maximal, in the sense, that black holes have more entropy than other objects with the same volume \cite{3,4}.
This maximum entropy of a black hole $S_0$, is related to its area $A$, as   $S_0 ={A}/{4}$. So, the  maximum entropy of a region   scales with the area of the boundary enclosing that region \cite{5}. This scaling behavior of the maximum  entropy has motivated the  development of the holographic principle \cite{6a, 7a}. It may be noted that at very small scales (close to the Planck scale), the holographic principle is expected to get violated due to quantum gravitational effects \cite{8,9}. So, the quantum gravitational effects are also expected to modify this entropy-area relation, at very small scales. The effect of quantum fluctuations, can be calculated from a short distance correction to the area-entropy law. This is because, the geometry of space-time can be obtained from thermodynamics in the Jacobson formalism \cite{25a,25b}. So, thermal fluctuations in the thermodynamics would correspond to quantum fluctuations in the geometry \cite{annals}. Furthermore, these quantum fluctuations in the geometry can be neglected   for very large black holes. However, such large black holes have a very small temperature, and the effects of thermal fluctuations can also be neglected for such black holes. Now these black holes evaporate with Hawking radiation, and reduce in size. At sufficiently small size, the effects of quantum fluctuations cannot be neglected, and these fluctuations can be analyzed perturbation. At such a small size, these black holes will also have a large temperature, and the effects of thermal fluctuations can also not be neglected at this stage. So, at this stage the thermal fluctuations can be analyzed as perturbations around the equilibrium thermodynamics \cite{PRD}.
It may be noted that as the black hole keeps reducing in size and it's the temperature keeps increasing, a stage is reached at which the manifold description of space-time breaks down, and at this stage the equilibrium description of thermodynamics also breaks down.   So, the correction to the equilibrium thermodynamics is only relevant at an intermediate scale, such that the scale is not so small that the equilibrium description breaks down, and it is not so large that effects by fluctuations can be neglected. In this paper, we shall analyze our system at such an    intermediate scale, where the thermal fluctuations can be expressed as perturbations around equilibrium thermodynamics.\\
We will analyze the effects of  fluctuations on a black hole in   massive  gravity. The massive gravity is constructed using the Vainshtein mechanism \cite{40,50}. In this mechanism, the general relativity is recovered in suitable limits \cite{50}. This  theory based on the Vainshtein mechanism contains  higher derivative terms, and these higher derivative terms  produce  negative norm   Boulware-Deser ghosts \cite{60}. It is possible to  resolve this problem for a subclass of massive potentials,  as  these Boulware-Deser ghosts do not exist for such a  subclass
\cite{70}-\cite{130}. This is  done by using the dRGT mechanism, and in this mechanism,  there  is a fixed metric along with the dynamical metric \cite{140}.\\
An asymptotically A(dS) charged BTZ black hole in a massive theory of gravity has also been constructed \cite{main}.
We will analyze the effects of quantum fluctuations on this BTZ black hole. These quantum fluctuations can be related to thermal
fluctuations in the thermodynamics of this BTZ black hole. It may be noted that effects of thermal fluctuations on the
thermodynamics of black hole has been studied, and it has been observed that such thermal fluctuations correct the entropy of
black holes \cite{bh12, bh14}. Such thermal fluctuations have also been studied for a small singly spinning Kerr-AdS black hole ~\cite{mir1}.
The effect of thermal fluctuations on the phase transitions in a sufficiently small AdS black holes has also been studied~\cite{mir2}.
Such thermal fluctuations have also been studied for different black saturns \cite{bs12, bs14}.   The effects of thermal fluctuations on
the thermodynamics of modified Hayward black hole have also been discussed, and phase transition for this system has also been investigated
\cite{u1}.\\
Now we can start from a three dimensional massive theory of gravity \cite{140}. The action for such a theory of massive gravity,   can be expressed as \cite{main},
\begin{equation}\label{action}
I=-\frac{1}{16\pi}\int d^{3}x \sqrt{-g}\left[\mathcal{R}-2\Lambda +L(\mathcal{F})+m^{2}\sum_{i=1}^{4} c_i{\cal U}_i(g, f)\right],
\end{equation}
where $\mathcal{R}$ is the scalar curvature,   $L(\mathcal{F})$ is the Lagrangian for the electromagnetic gauge field, and  $\Lambda$ is the cosmological constant. Here $m$ is the mass term and  $f$ is the reference metric. Furthermore,  the  suitable constants for massive gravity are denoted by   $c_i$, and the symmetric
polynomials of the eigenvalues of the $d\times d$
matrix ${\mathcal{K}^{\mu}}_{\nu}=\sqrt{g^{\mu\alpha}f_{\alpha\nu}} $ are denoted by  $\mathcal{U}_i$, such that
\begin{eqnarray}\label{Ui}
&&\mathcal{U}_{1}=[\mathcal{K}],\nonumber\\
&&\mathcal{U}_{2}=[\mathcal{K}]^2 -[\mathcal{K}^2],\nonumber\\
&&\mathcal{U}_{3}=[\mathcal{K}]^3 -3[\mathcal{K}][\mathcal{K}^2]  + 2[\mathcal{K}^3],\nonumber\\
&&\mathcal{U}_{4}=[\mathcal{K}]^4 -6[\mathcal{K}^2][\mathcal{K}]^2  + 8[\mathcal{K}^3][\mathcal{K}]+3[\mathcal{K}^2]^2
-6[\mathcal{K}^4].
\end{eqnarray}
We also have ${(\sqrt{A})^{\mu}}_{\nu}{(\sqrt{A})^{\nu}}_{\lambda}={A^{\mu}}_{\lambda}$ and $\mathcal{K}={\mathcal{K}^{\mu}}_{\mu}$.
Now a static charged black hole with(A)dS asymptotes is given by \cite{main},
\begin{equation}\label{metric1}
ds^2= -f(r)dt^2 +\frac{dr^2}{f(r)}+r^2 d\varphi^2,
\end{equation}
where $f(r)$ can be expressed as
\begin{equation}
f(r)=-\Lambda r^2-m_{0}-2q^{2}\ln{(\frac{r}{l})}+m^{2}c c_{1} r,
\end{equation}
here $m_{0}$ is a constant related to the black hole mass ($M$) as $m_{0}=8M$, and $q$
is another constant related to the black hole electric charge ($Q$) as $q=2Q$. Also, $m$ is a massive parameter,
while $c, c_{1}$ are positive constants. In the case of $m=0$, we have the linearly charged BTZ black hole solution.
We should note that asymptotically behavior of the solution is dS for $\Lambda>0$, while AdS for $\Lambda<0$. It can be
argued that the structure of the horizon   dependent on these black hole parameters. Hence, choosing suitable parameters, we obtain
two event horizons $r_{+}$ and $r_{-}$. It may be noted that for  dS,
we need to choose suitable parameters, such the temperature at
the black hole horizon is the same as the cosmological horizon.
It order to do that we need to analyze the structure of the  horizon.
This can be done by using   the approximate  ${r}/{l}\ll1$ and   $\ln{1+x}\approx x$, to obtain
\begin{equation}
r_{\pm}\approx{\frac { \left( c{m}^{2}c_{1}l-4\,{q}^{2}\pm A \right) l}{2({l}^{2}\Lambda-{q}^{2})}},
\end{equation}
where we have defined $A$ as
\begin{equation}\label{A}
 A = \sqrt {{c}^{2}{m}^{4}{c_{1}}^{2
}{l}^{2}-8\,c{m}^{2}c_{1}l{q}^{2}+12\,{l}^{2}\Lambda\,{q}^{2}-4\,{l}^{2}
\Lambda\,m_{0}+4\,{q}^{4}+4\,m_{0}{q}^{2}}.
\end{equation}
We will discuss the specific values of this parameter after introducing temperature.
At such specific values,
 we will have thermodynamic equilibrium. Thus, we obtain,
\begin{equation}
M=-\frac{\Lambda}{8} r_{+}^2-Q^{2}\ln{(\frac{r_{+}}{l})}+\frac{m^{2}c c_{1}}{8} r_{+}.
\end{equation}
The Hawking temperature is given by,
\begin{equation}\label{Temp1}
T_{\pm}=-\frac{\Lambda r_{\pm}}{2\pi}-\frac{q^{2}}{2\pi r_{\pm}}+\frac{m^{2}c c_{1}}{4\pi}.
\end{equation}
It should be noted that the temperature at both horizons must be the same, and this yields to the following condition,
\begin{eqnarray}
  \sqrt {4\,{q}^{4}+ \left( -8\,c{m}^{2}c_{1}l+12\,{l}^{2}\Lambda+4
\,m_{0} \right) {q}^{2}+{l}^{2} \left( {c}^{2}{m}^{4}{c_{1}}^{2}-4\,\Lambda\,m_{0}
 \right) } \nonumber \\\times 2\, \left( \Lambda\, \left( -4\,{q}^{2}+m_{0} \right) {l}^{2}+{q}^{4}
 \right)=0.
\end{eqnarray}
It give us two conditions to have the same temperatures at both the horizons,
\begin{eqnarray}
\left( \Lambda\, \left( -4\,{q}^{2}+m_{0} \right) {l}^{2}+{q}^{4}
 \right)&=&0,\\
{4\,{q}^{4}+ \left( -8\,c{m}^{2}c_{1}l+12\,{l}^{2}\Lambda+4
\,m_{0} \right) {q}^{2}+{l}^{2} \left( {c}^{2}{m}^{4}{c_{1}}^{2}-4\,\Lambda\,m_{0}
 \right) }&=&0.
\end{eqnarray}
It should be noted that  by setting  $A=0$ in the equation (\ref{A}) we get the  extremal limit for the model, and this is exactly the second condition above.
Now from  first condition, we obtain,
\begin{equation}\label{condition11}
\Lambda=\frac{q^{4}}{(4q^{2}-m_{0})l^{2}}.
\end{equation}
In the dS space with positive $\Lambda$,  we must have $4q^{2}>m_{0}$. Using the second condition, we obtain
\begin{equation}\label{condition22}
\Lambda={\frac {{c}^{2}{m}^{4}{c_{1}}^{2}{l}^{2}-8\,c{m}^{2}c_{1}l{q}^{2}
+4\,{q}^{4}+4\,m_{0}{q}^{2}}{4{l}^{2} \left( -3\,{q}^{2}+m_{0} \right) }}.
\end{equation}
Combining two above equations, we obtain the following  condition
\begin{eqnarray}\label{condition111}
-4\,{c}^{2}{m}^{4}{c}^{2}{l}^{2}{q}^{2}+{c}^{2}{m}^{4}{c_{1}}^{2}{l}^{2}m_{0}+
32\,c{m}^{2}c_{1}l{q}^{4}&&\nonumber\\ -8\,c{m}^{2}c_{1}lm_{0}{q}^{2}-28\,{q}^{6}-8\,m_{0}{q}^{4}+4
\,{m_{0}}^{2}{q}^{2}&=&0.
\end{eqnarray}
It can be used to fix parameters of the model,
for example if we choose $m=m_{0}=l=c=c_{1}=1$, then one can find $q=0.55$.
It may be noted that there are other possibilities which can be used to constraint the parameter space.
So,   we will also use both conditions (\ref{condition11}) and (\ref{condition22})
to study black hole entropy which help us to further  constraint the parameter ($q$).

\begin{figure}[h!]
 \begin{center}$
 \begin{array}{cccc}
\includegraphics[width=70 mm]{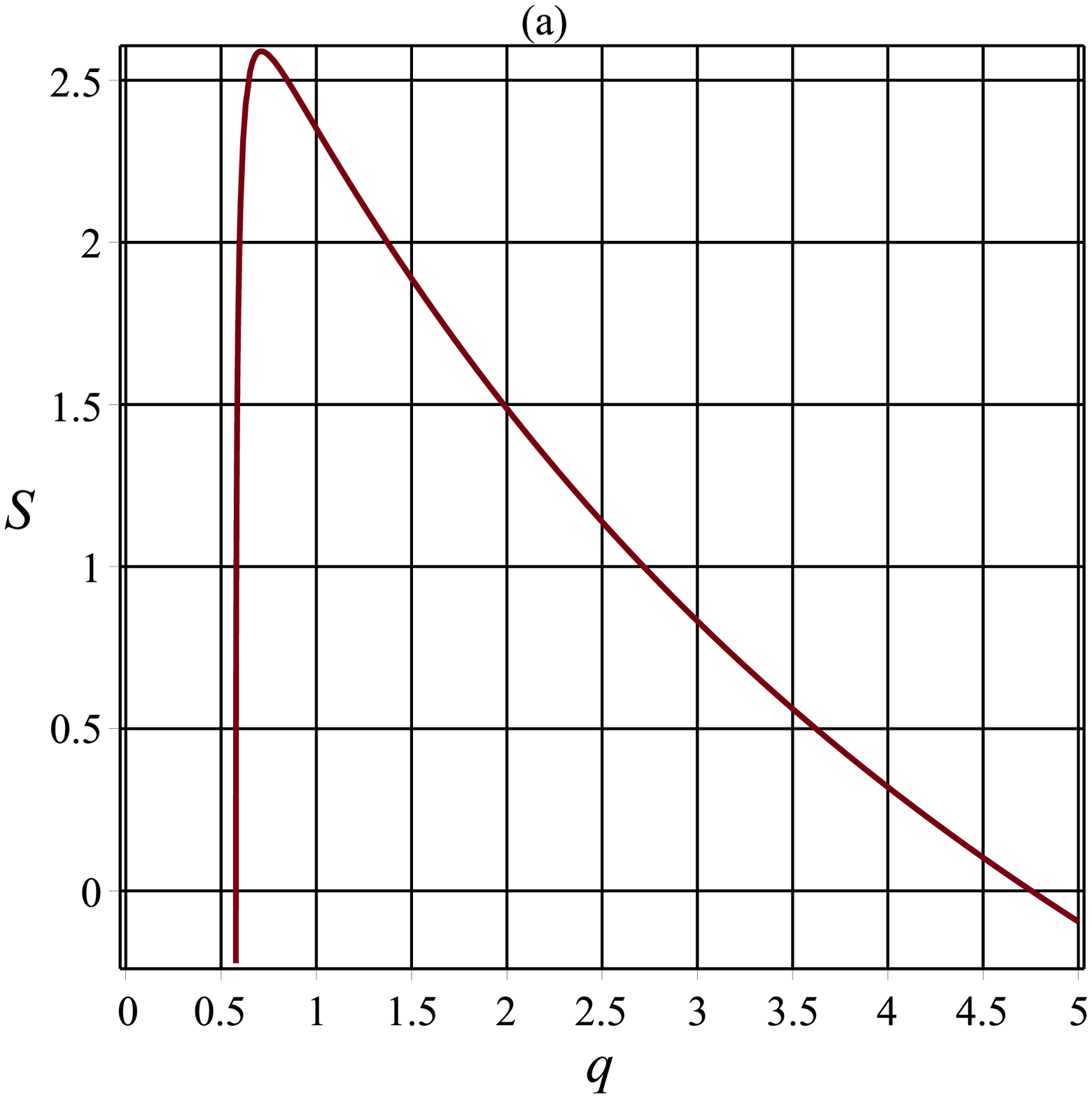}\includegraphics[width=70 mm]{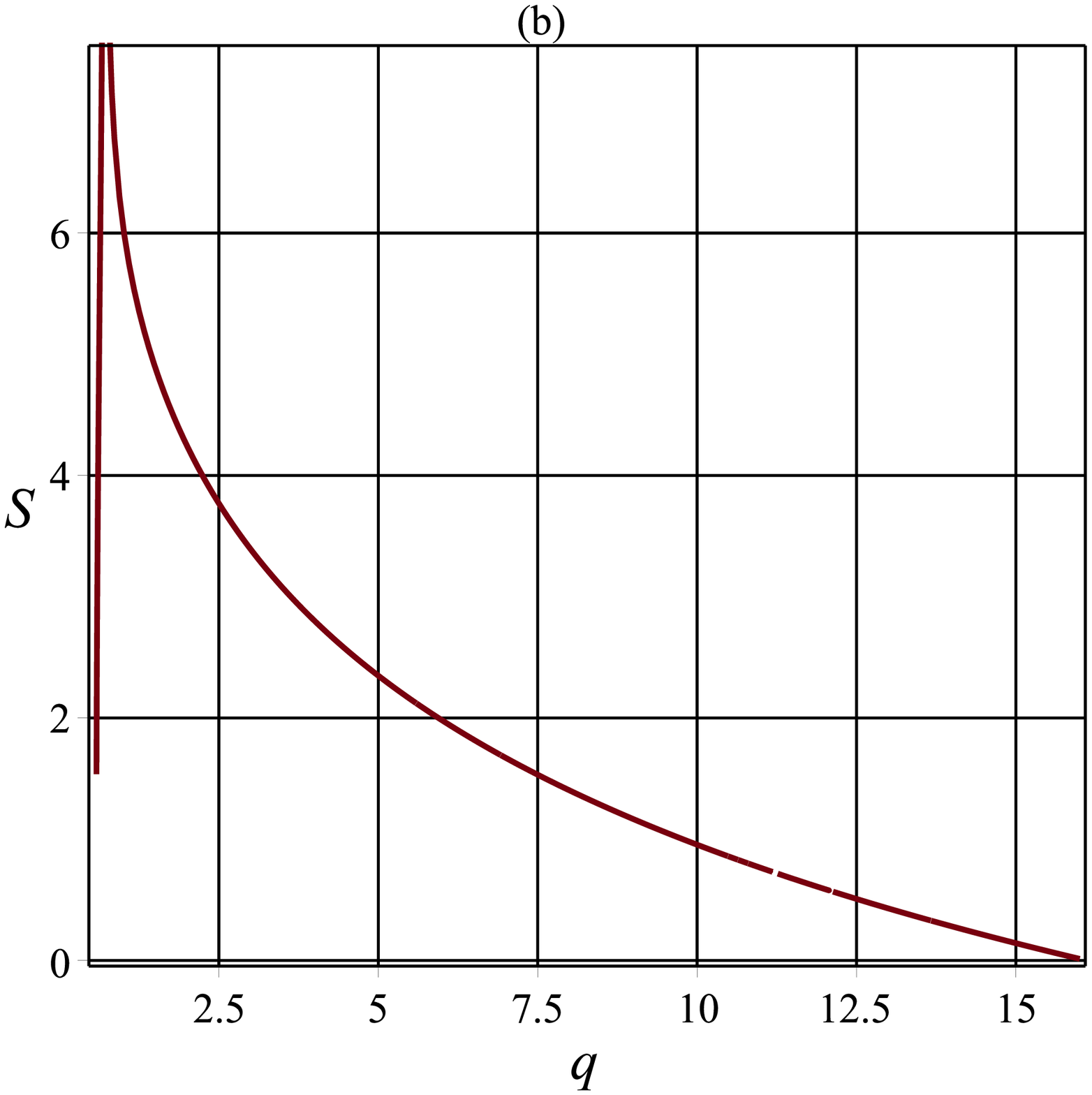}
 \end{array}$
 \end{center}
\caption{Corrected entropy in terms of $q$. We have set
unit values for all other parameters. (a) By using the condition (\ref{condition11}).
(b) Extremal case given by the  condition (\ref{condition22}).}
 \label{fig0}
\end{figure}

\begin{figure}[h!]
 \begin{center}$
 \begin{array}{cccc}
\includegraphics[width=65 mm]{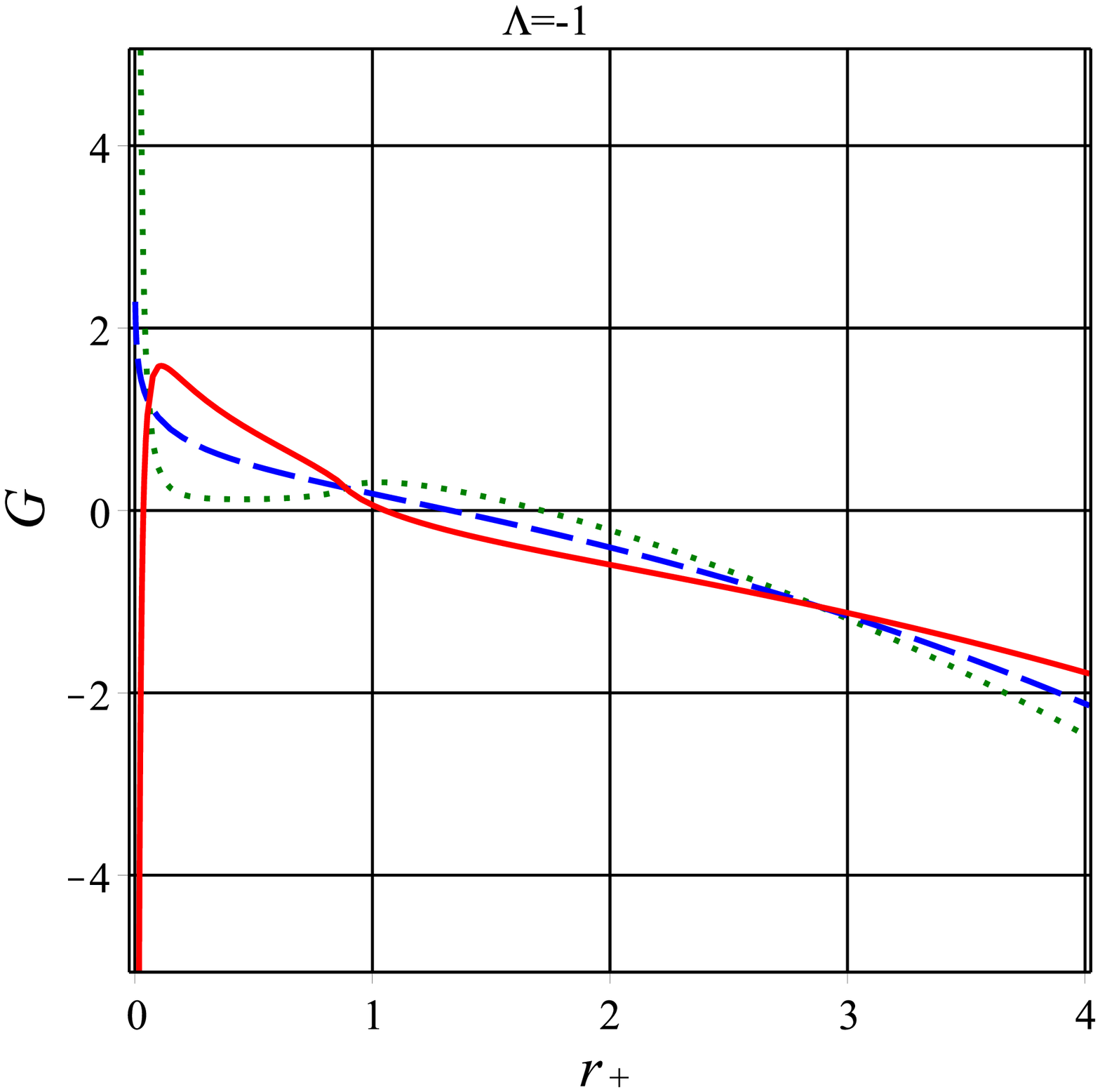}
\includegraphics[width=65 mm]{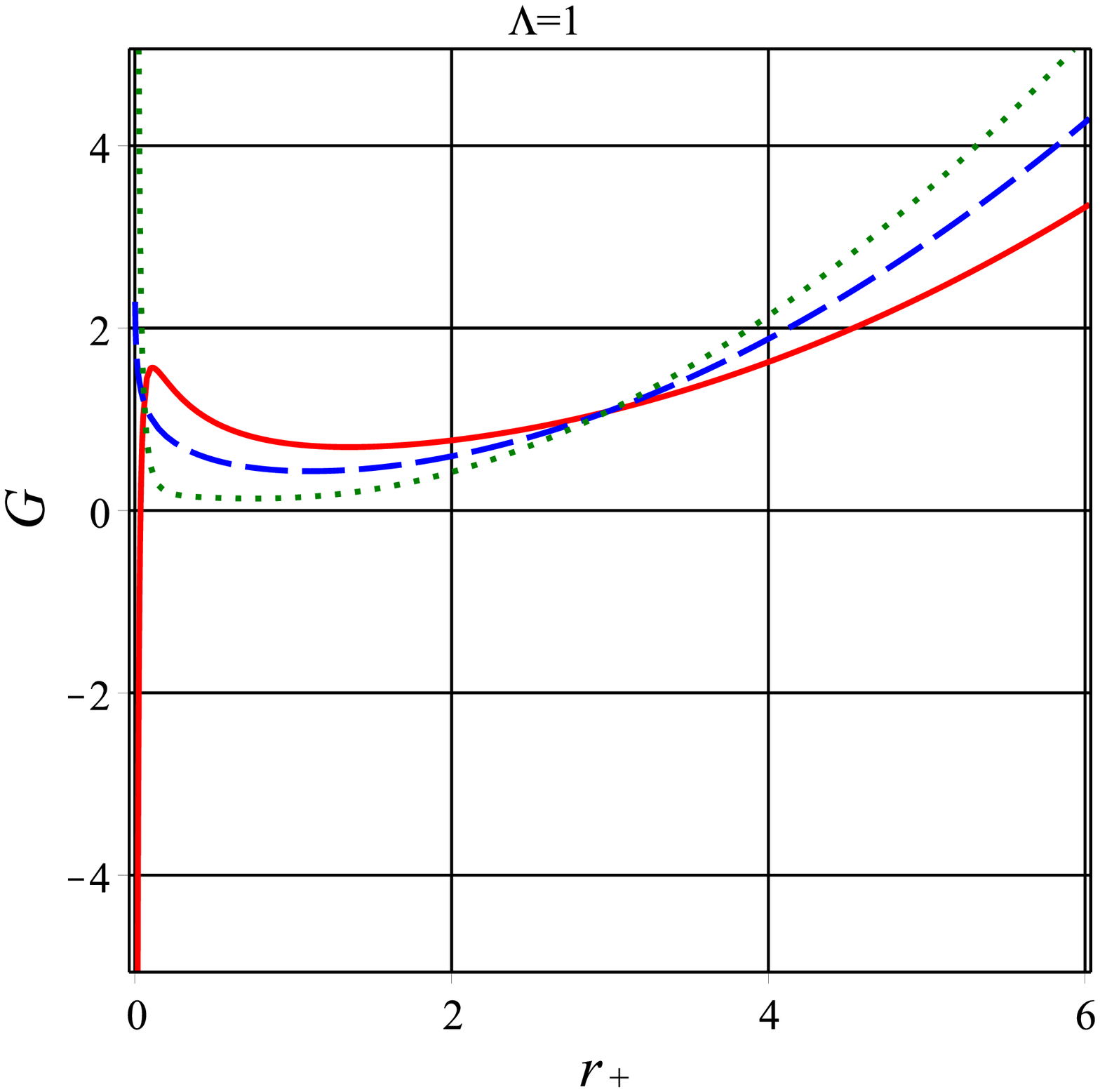}
 \end{array}$
 \end{center}
\caption{Gibbs free energy in terms of $r_{+}$, for $q=0.55$ and
unit values for all other parameters. The blue dashed lines represent the case of $\alpha=0$,
while the red solid lines represent the case of $\alpha=1$, and green dotted lines represent the case of $\alpha=-1$.}
 \label{fig1}
\end{figure}

We can now analyze the corrections to the black hole entropy of the BTZ black hole due to quantum  fluctuations.
It may be noted  the original black hole entropy  for the BTZ black hole, in massive gravity, can be written as
\begin{equation}\label{Ent1}
S_{0}=\frac{\pi}{2}r_{+}.
\end{equation}
We can calculate the corrections to this original black hole entropy using the path integral quantization of gravity.
The path integral quantization of gravity can be done using the Euclidean quantum gravity \cite{o, 01ab, 02ab, 04ab, 05ab}.
In this formalism,
the temporal coordinate is rotated in  the complex plane,
and the quantum gravitational partition function resembles a statistical mechanical partition
function \cite{hawk,hawk1}.
Thus, if $I_E=-i I$ is the Euclidean action obtain from action given by Eq. (\ref{action}),
then we can write
\begin{equation}
Z = \int [D]\exp(-I_E).
\end{equation}
This can be written as a  statistical mechanical partition function, with $\beta$
as the inverse temperature \cite{hawk,hawk1},
\begin{equation}
Z = \int_0^\infty  dE\rho (E)\exp(-\beta E).
\end{equation}
The density of states corresponding to this statistical mechanical partition function can be
written as
\begin{eqnarray}
\rho(E) =\frac{1}{2 \pi i} \int^{\beta_0+ i\infty}_{\beta_0 -i\infty}d \beta\exp[S(\beta)].
\end{eqnarray}
Here the entropy $S$ can be expressed as
\begin{equation}
S = \beta  E + \ln Z.
\end{equation}
This system is usually studied at the equilibrium
temperature $\beta_0$. Thus, for this system usually the black hole thermodynamics
can be obtained by neglecting all the thermal fluctuations. So, using
$T=\beta^{-1}$, it is possible to obtain $S_0 = S(\beta)_{\beta = \beta_0} =A/ 4G_N$,
where $A$  is the black hole area \cite{th}. The quantum fluctuations increase with the decrease in the size of the black hole. This corresponds to an
increase in the thermal fluctuations, in this statistical mechanical partition function.
At Planck scale the full equilibrium description of thermodynamics cannot be applied,
as the space-time cannot be analyzed as a differential manifold. However, in an intermediate
regime, where the black hole is not small enough for the equilibrium description to break down and not large enough for the quantum  fluctuations to have no effect, these   fluctuations can be studied as thermal fluctuations in  the partition
function of Euclidean quantum gravity. The effects of thermal fluctuations on thermodynamics of a black hole are known \cite{bh12, bh14},  so
we can obtain  the correct entropy $S(\beta)$, at a temperature $\beta$ as a perturbation around the equilibrium temperature $\beta_0$
\begin{equation}
S = S_0+\frac{1}{2}(\beta-\beta_0)^2
\left(\frac{\partial^2 S(\beta)}{\partial \beta^2 }\right)_{\beta =\beta_0}.
\label{a1}
\end{equation}
Here the effects of higher order corrections have been neglected.
Now we can define $S_0 = S(\beta)|_{\beta = \beta_0}$, and express the
density of states as
\begin{eqnarray}
\rho(E)&=& \frac{\exp(S_0)}{2 \pi i}\int^{\beta_0+i\infty}_{\beta_0-i\infty}d\beta\nonumber \\&&
\times
\exp \left( \frac{1}{2}(\beta-\beta_0)^2 \left(\frac{\partial^2 S(\beta)}{\partial \beta^2 }
\right)_{\beta =\beta_0}\right).
\end{eqnarray}
After performing a change of variables, this can be expressed as
\begin{equation}
\rho(E) = \frac{\exp(S_{0})}{\sqrt{2\pi}} \left[\left(\frac{\partial^2
S(\beta)}{\partial \beta^2 }\right)_{\beta = \beta_0}\right]^{- 1/2},
\end{equation}
Thus, we can write
\begin{equation}
S = S_0 -\frac{1}{2}
\ln \left[\left(\frac{\partial^2 S(\beta)}{\partial \beta^2 }\right)_{\beta= \beta_0}\right].
\end{equation}
We can use the microscopic degrees of freedom obtained
from a conformal field theory to analyze this corrected entropy.
The modular invariance of such a conformal field theory can be used to argue that the    entropy  should be expressed as
$S(\beta) = a \beta+b\beta^{-1}$~\cite{card},  and this has been   generalized to
$S(\beta) = a \beta^m+ b \beta^{-n}$~\cite{bh12},
where $m, n, a, b $ are positive constants.
This expression has an extremum at $\beta_0 = (nb/ma)^{1/ m+n} = T^{-1}$, and  the corrected entropy around this extremum can be expressed as,
\begin{eqnarray}
S(\beta) &=& [(n/m)^{m/(m+n)} + (m/n)^{n/(m+n)} ](a^n b^m)^{1/(m+n)}
\nonumber \\
&&
+\frac{1}{2}[(m+n) m^{(n+2)/(m+n)} n^{(m-2)/(m+n)}]
\nonumber \\ && \times
(a^{n+2}b^{m-2})^{{1}/(m+n)}(\beta - \beta_0)^2.\label{a2}
\end{eqnarray}
Thus, we obtain the expression for the corrected entropy as
\begin{equation}
S  = (n/m)^{m/(m+n)} + (m/n)^{n/(m+n)} (a^n b^m)^{{1}/(m+n)}.
\end{equation}
So, finally, we obtain
\begin{eqnarray}
\left(\frac{\partial^2 S(\beta)}{\partial \beta^2 }\right)_{\beta = \beta_0} =  \mathcal{Y}_{mn} S_0 T^2~,
\end{eqnarray}
where $\mathcal{Y}_{mn}$ is a factor which depends on $m$ and, $n$, hence they can be absorbed by a suitable redefinition, as it does not depend on any parameter of this BTZ black hole. The corrections to the entropy of this BTZ black hole scale as $\ln S_0 T^2$ \cite{bh12,bh14}. Now it is known  that such logarithmic corrections to entropy of a black hole are obtained in almost all approaches to quantum gravity, and this seems to be a universal model independent result. In fact, such logarithmic corrections have been obtained using non-perturbative quantum general relativity ~\cite{10a}. Hence, logarithmic term may reflect quantum corrections \cite{Mann2}. String theoretical corrections to black hole entropy have also been calculated, and it has been observed that such correction terms can be expressed  using logarithmic functions~\cite{15a,16a,17a,18a}. Using the Cardy formula, it has been argued that black holes  whose microscopic degrees of freedom are governed by a conformal field theory must produce logarithmic corrections to the area-entropy relation~\cite{11, 12}. It may be noted that even though the form of these corrections seems to be a model independent result, the coefficient of this logarithmic correction terms depends on the details of the model, and is a model dependent result.
So, we will keep this coefficient as a general parameter $\alpha$, and write the corrected entropy as \cite{Rahim},
\begin{equation}\label{Corrected-Entropy}
S=S_{0}-\frac{\alpha}{2}\ln{(S_{0}T^{2})},
\end{equation}
where correction parameter $\alpha$ added by hand to find effect of such correction clearly in analytical expressions.\\
Thus, we can write the
 logarithmic corrected entropy, for the BTZ black hole, in massive gravity as
\begin{equation}\label{Corrected-Entropy1}
S=\frac{\pi}{2}r_{+}-\frac{\alpha}{2}\ln{\left[\frac{r_{+}}{8\pi}\left(-\Lambda r_{+}-\frac{q^{2}}{r_{+}}+\frac{m^{2}c c_{1}}{2}\right)^{2}\right]}.
\end{equation}
We should use conditions (\ref{condition11}) and (\ref{condition22}) to draw corrected entropy and see that is positive. According to the Eq.  (\ref{condition111}),
by choosing unit value for all parameter instead $q$,
we have $q\approx0.55$, so we can draw behavior of entropy in terms of $q$ in the Fig. \ref{fig0} (a)
for the condition (\ref{condition11}) and (b) for the extremal case with the equation (\ref{condition22}).
We observe that entropy is positive for $q>0.55$. However very large value of the black hole charge  may
also yields a negative entropy,  hence we are restricted to choose $q$ in the finite interval. From the Fig. \ref{fig0} (a)
we can see restricted interval is $0.55\leq q\leq 4.75$, while for the extremal case, as illustrated by the Fig. \ref{fig0}
(b) we can find $0.55\leq q\leq 16$. It means that the extremal case yields to larger range for the electrical charge of black hole.\\
The first law of black hole thermodynamics is given by,
\begin{equation}\label{FirstL1}
dM=TdS+\Phi dQ,
\end{equation}
where $\Phi$ is the electric potential, and it is given by \cite{main},
\begin{equation}\label{U1}
\Phi=-2Q\ln{(\frac{r_{+}}{l})}.
\end{equation}
The first law (\ref{FirstL1}) is valid for the case  $\alpha=0$. In presence of $\alpha$, the first law is valid if the
following condition is satisfied
\begin{equation}\label{condition1}
6\Lambda r_{+}^{2}-m^{2}c c_{1} r_{+}-8Q^{2}=0.
\end{equation}
Thus, the   horizon radius can be expressed as
\begin{equation}\label{horizon1}
r^{\ast}_{+}=\frac{m^{2}c c_{1}}{12\Lambda}\left(1+\sqrt{1+\frac{192\Lambda q^{2}}{m^{4}c^{2} c_{1}^{2}}}\right).
\end{equation}
This means that in presence of logarithmic correction, the first law of black hole thermodynamics is valid only at $r_{+}=r^{\ast}_{+}$.\\
In this case, the internal energy given by,
\begin{equation}\label{Int-E1}
E=\int{TdS},
\end{equation}
which yields the following expression,
\begin{equation}\label{Int-E1-1}
E=\frac{m^{2}c c_{1}}{8}r_{+}-\frac{\Lambda}{8}r_{+}-Q^{2}\ln{r_{+}}+\frac{\alpha}{\pi}\left[\frac{5}{4}r_{+}+\frac{Q^{2}}{r_{+}}-\frac{m^{2}c c_{1}}{8}\ln{r_{+}}\right].
\end{equation}
We find that the internal energy is decreasing function of $\alpha$ in the AdS space-time, while it is an increasing function of $\alpha$ in dS space-time. It has similar behavior with the case of $m=0$. In both (A)dS space-times,  it is increasing function of massive parameter $m$, and  a decreasing function of the black hole charge $Q$. Then we can study Gibbs free energy using the following relation,
\begin{equation}\label{Gibbs1}
G=M-TS.
\end{equation}
In the plots of the Fig. \ref{fig1}, we can see typical behavior of the Gibbs free energy in terms of $r_{+}$.
This can be done by varying    $\alpha$, and observing  the effects of the logarithmic correction. As we expected, the main difference occurs only  at small $r_{+}$. In presence of thermal fluctuations with positive coefficient, value of the Gibbs free energy is negative infinite, and it is positive infinite for $\alpha=-1$ if we do not  consider thermal fluctuations.  This result holds for both AdS and dS space-times.\\

\begin{figure}[h!]
 \begin{center}$
 \begin{array}{cccc}
\includegraphics[width=65 mm]{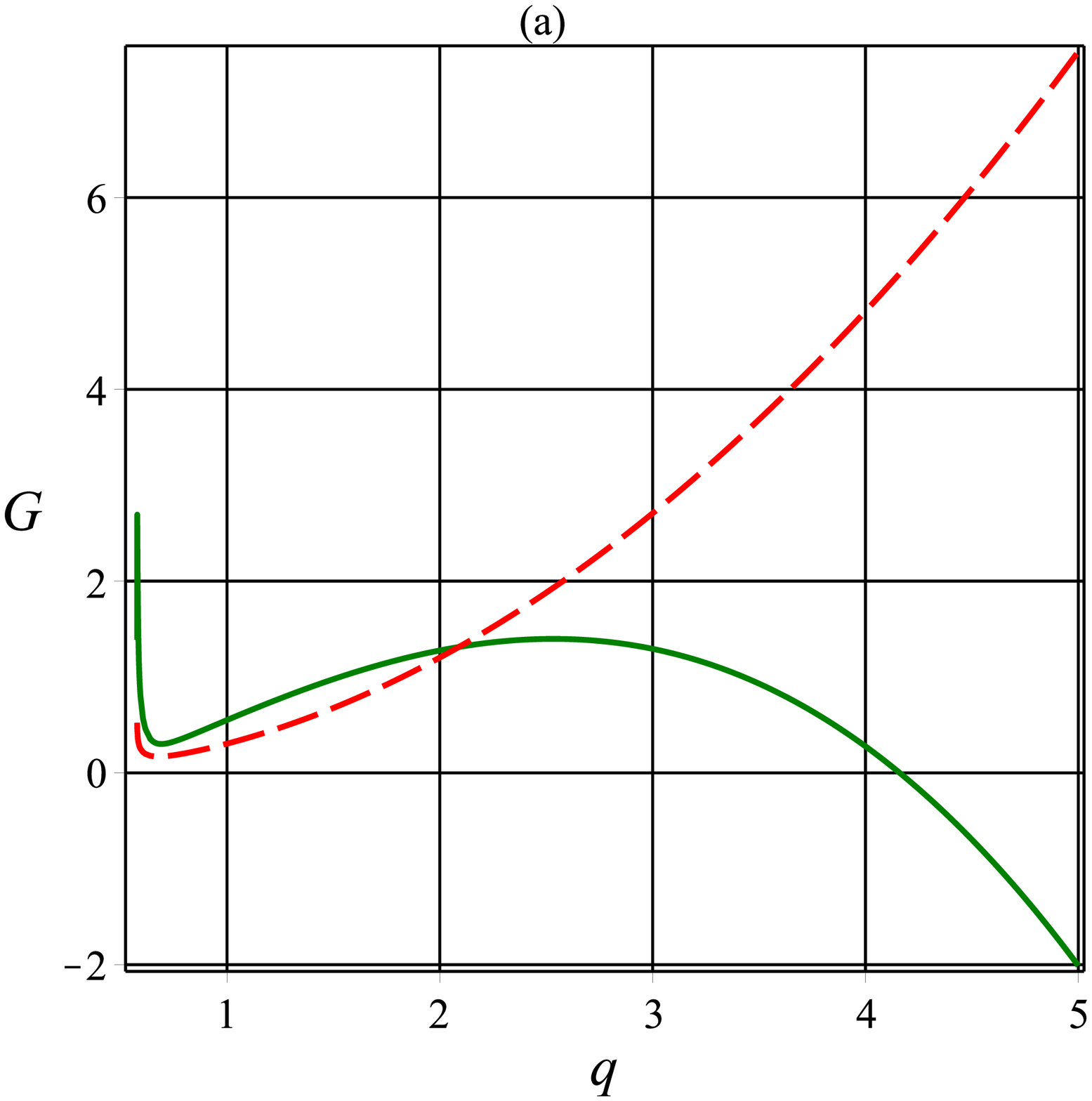}
\includegraphics[width=65 mm]{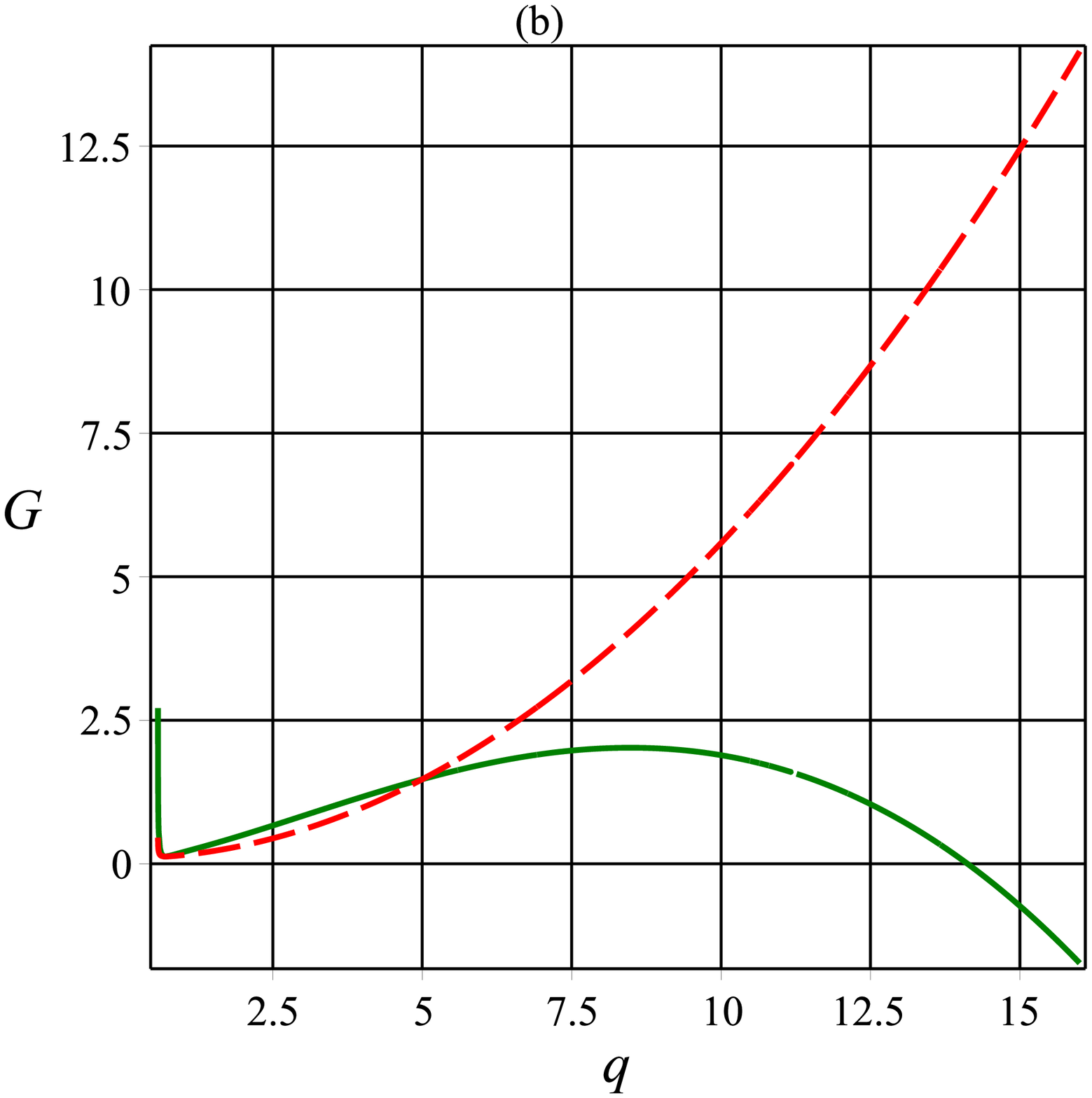}
 \end{array}$
 \end{center}
\caption{Gibbs free energy in terms of $q$, for unit values for all parameters. The red dashed lines represent the case of $\alpha=0$,
while the green solid lines represent the case of $\alpha=1$. (a) By using the condition (\ref{condition11}).
(b) Extremal case given by the condition (\ref{condition22}).}
 \label{fig1-1}
\end{figure}

Also, we examined both possibilities given by conditions (\ref{condition11}) and (\ref{condition22}) which is illustrated by plots of the Fig. \ref{fig1-1}. We can see that general behavior of the Gibbs free energy is the same.

\begin{figure}[h!]
 \begin{center}$
 \begin{array}{cccc}
\includegraphics[width=65 mm]{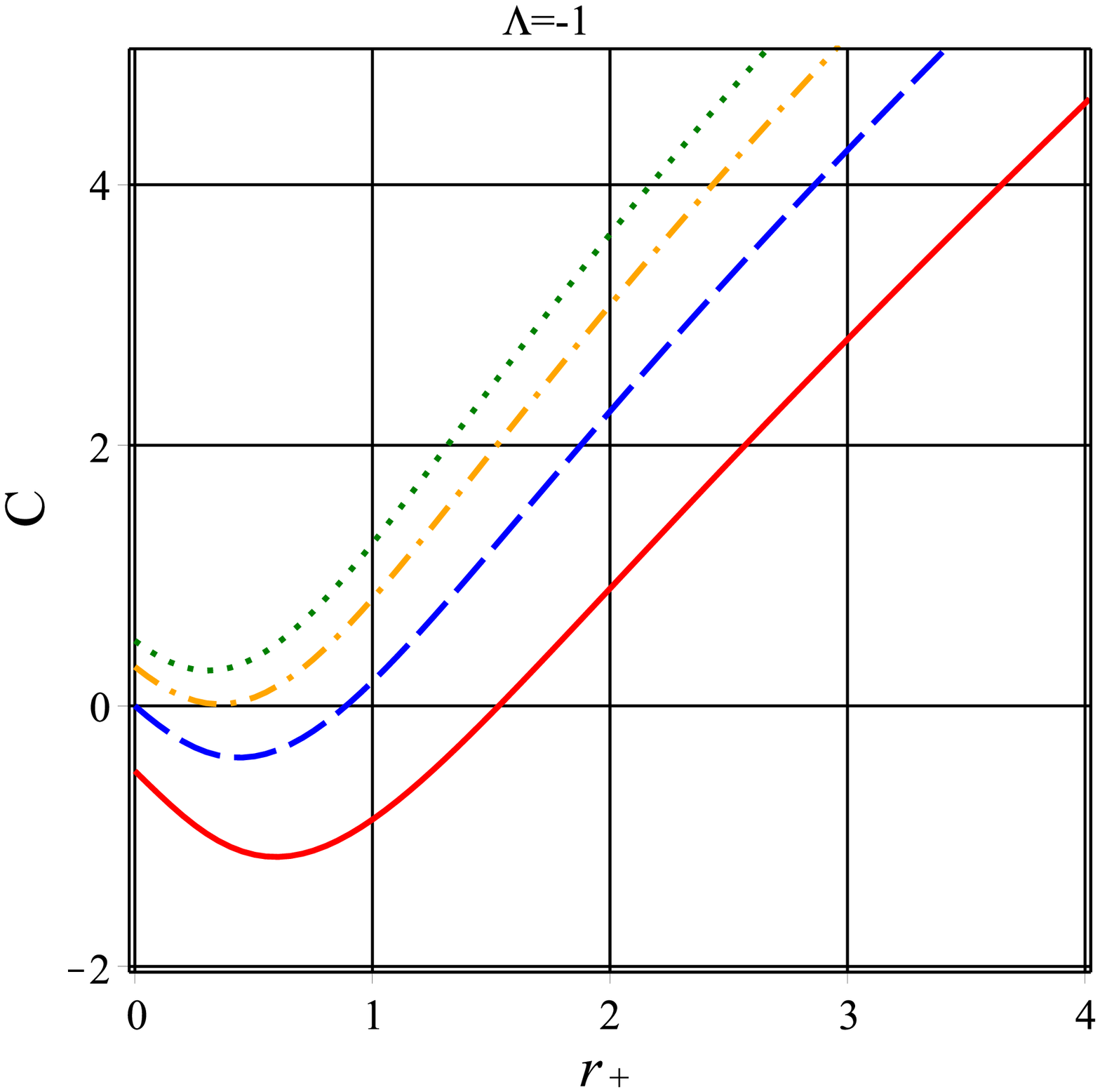}
\includegraphics[width=65 mm]{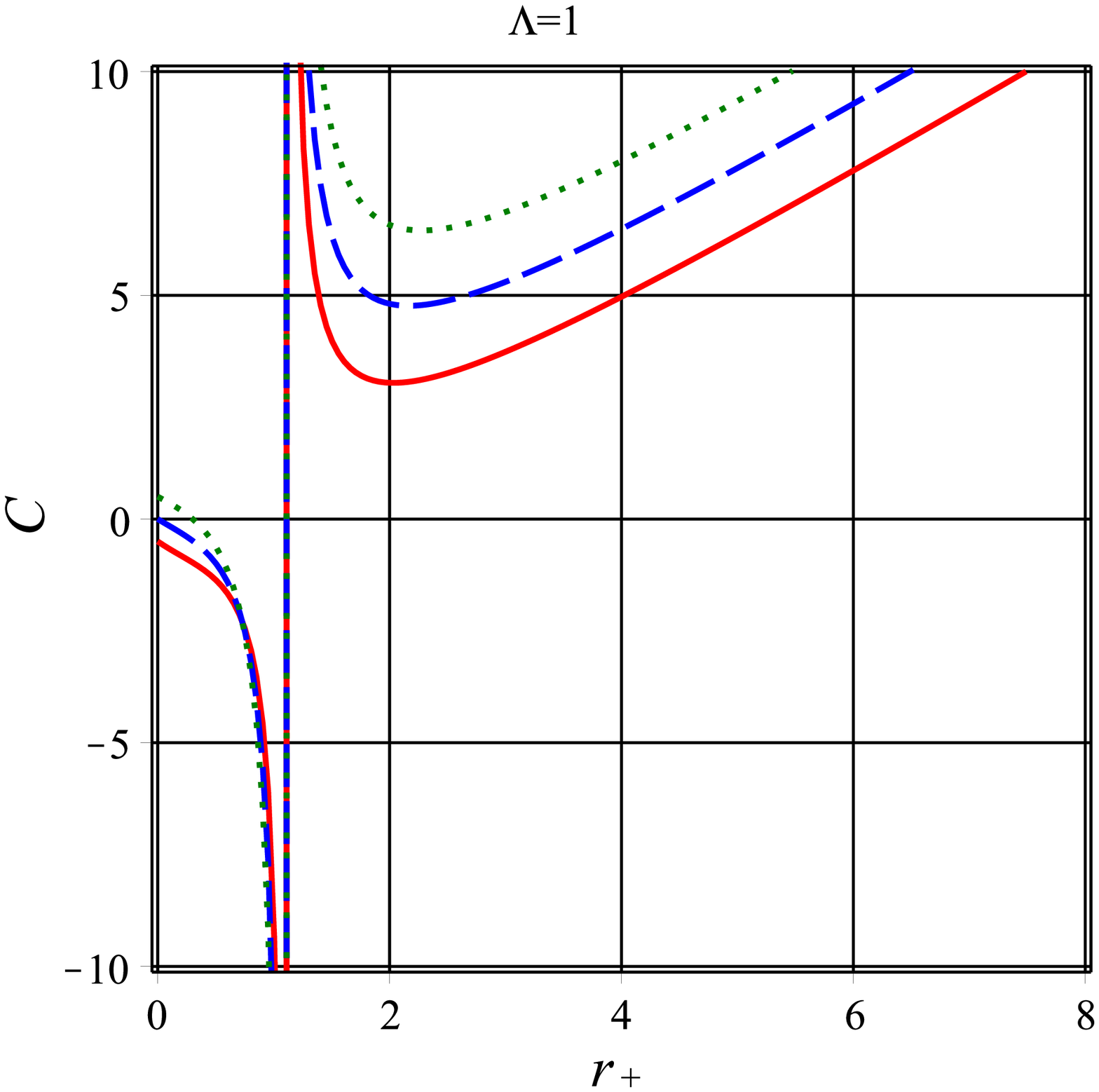}
 \end{array}$
 \end{center}
\caption{Specific heat in terms of $r_{+}$, for $q=0.55$ and
unit values for all  other parameters. The blue dashed lines represent the case of $\alpha=0$,
red solid lines represent the case of $\alpha=1$, and green dotted lines represent the case of $\alpha=-1$.
The orange dash dotted line represent the case of $\alpha=-0.6$.}
 \label{fig2}
\end{figure}

\begin{figure}[h!]
 \begin{center}$
 \begin{array}{cccc}
\includegraphics[width=65 mm]{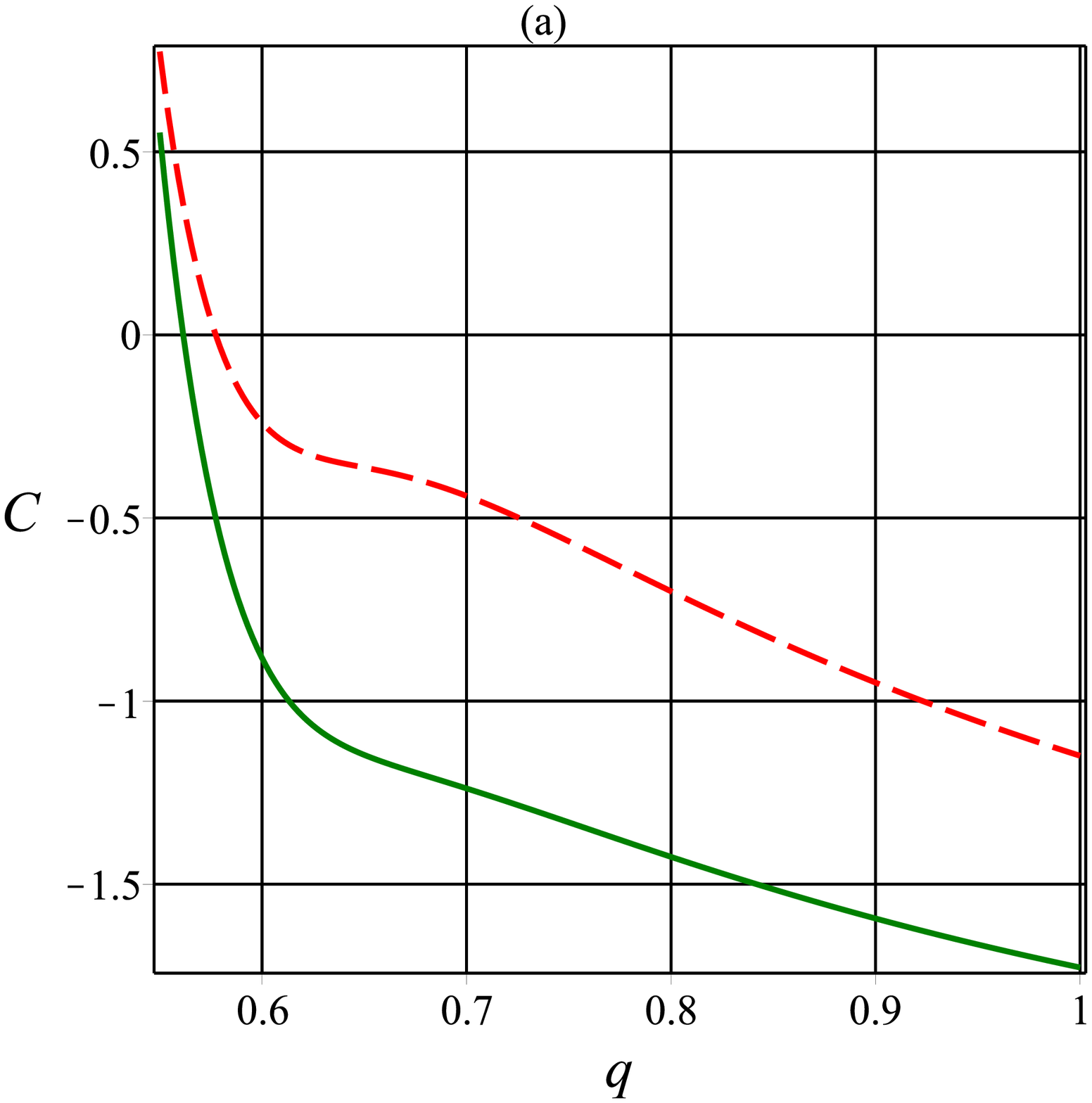}
\includegraphics[width=65 mm]{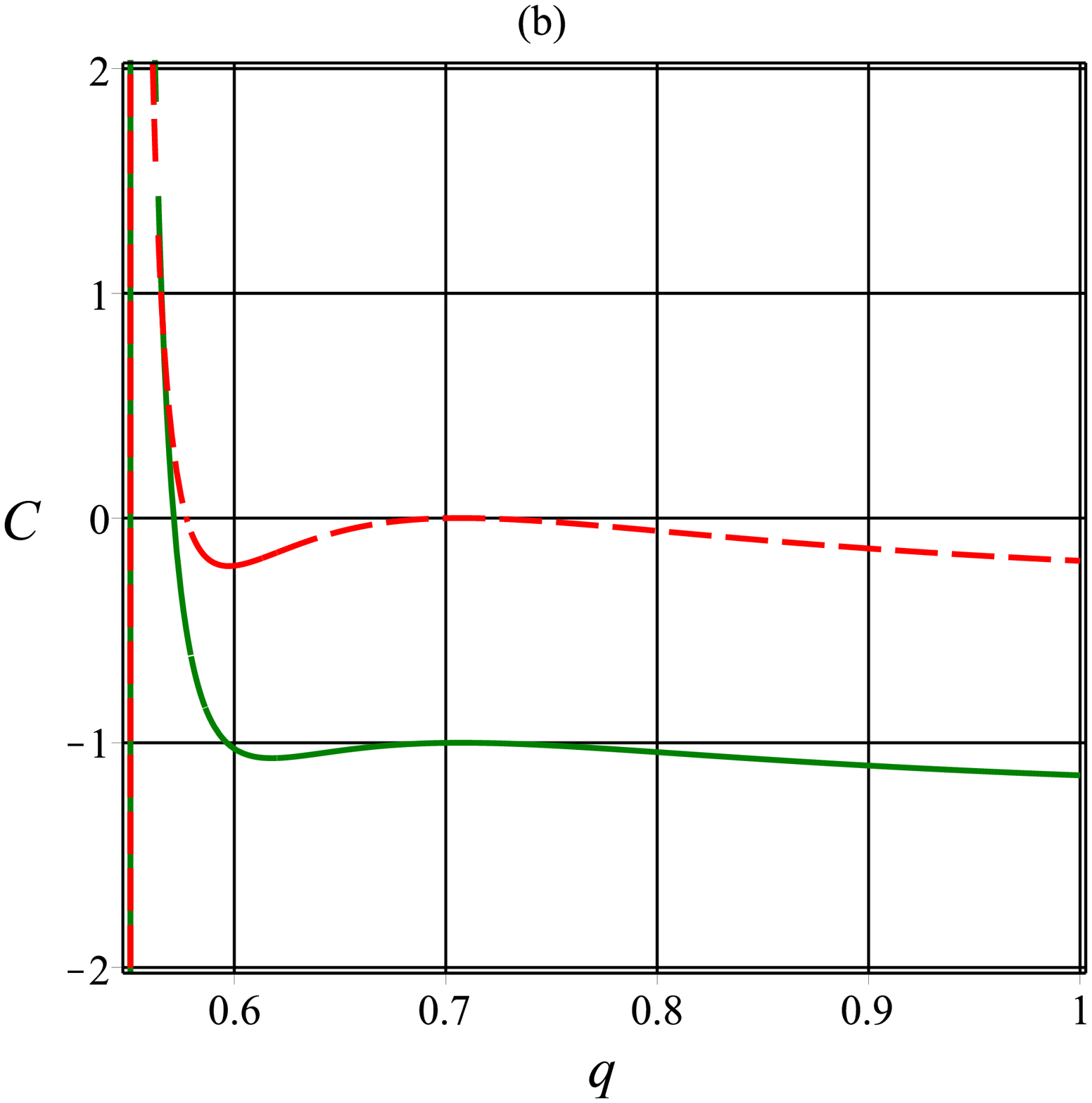}
 \end{array}$
 \end{center}
\caption{Specific heat in terms of $q$, for unit values for all parameters. The red dashed lines represent the case of $\alpha=0$,
while the green solid lines represent the case of $\alpha=1$. (a) By using  the condition (\ref{condition11}).
(b) Extremal case given by the condition (\ref{condition22}).}
 \label{fig2-1}
\end{figure}

There are some special points where corrected and uncorrected Gibbs free energy have the same value. In the case of AdS space-time there is a minimum for the Gibbs free energy which may be stable point.
In order to obtain more information about stable points we study sign of specific heat.\\
In the canonical ensemble, heat capacity at constant $Q$ is  calculated by using the following relation,
\begin{equation}\label{S-Heat1}
C=T(\frac{dS}{dT}).
\end{equation}
It yields to the following expression,
\begin{equation}\label{S-Heat1-1}
C=\frac{\pi}{4}\left(\frac{2\Lambda r_{+}^{2}-m^{2}cc_{1}r_{+}+8Q^{2}}{\Lambda r_{+}^{2}-4Q^{2}}\right)r_{+}
-\frac{\alpha}{4}\left(\frac{6\Lambda r_{+}^{2}-m^{2}cc_{1}r_{+}-8Q^{2}}{\Lambda r_{+}^{2}-4Q^{2}}\right).
\end{equation}
In the plots of Fig. \ref{fig2},  we can see behavior of specific heat for both AdS and dS space-times.
In the left plot of the Fig.\ref{fig2}, we can observe  behavior of  the specific heat in AdS space-time.
For  $\alpha=0$, we obtain  some negative regions, and these regions exhibit some instabilities.
It has already been observed by studding the equilibrium thermodynamics, and neglecting thermal fluctuations \cite{main}.
However, for   $\alpha=-1$ (for selected values of the black hole parameters), we have completely positive specific heat.
So,   the  logarithmic corrected entropy with negative coefficient can have a  stable   AdS space-time.
There is a special radius ($r_{+0}$), where $C=0$ which has been illustrated by orange dash dotted line
and is obtained as root of the following equation,
\begin{equation}\label{r+01}
\frac{\pi}{2}\Lambda r_{+0}^{3}-\left(\frac{\pi}{4}m^{2}cc_{1}+\frac{3}{2}\Lambda\alpha\right)r_{+0}^{2}+\left(2\pi Q^{2}+\frac{m^{2}cc_{1}}{4}\alpha\right) r_{+0}+2\alpha Q^{2}=0.
\end{equation}
For $r_{+}\geq r_{+0}$, the black hole is always in a stable phase, and for $r_{+}\leq r_{+0}$,
the stability of black hole  depends on value of $\alpha$. For the suitable negative value of $\alpha$, we have a stable
black hole for all $r_{+}$. In the case of AdS space-time, there is no phase transition which corresponds to divergence of specific heat.
On the other hand, in the dS space-time the phase transition occurs,  as is illustrated by right plot of the Fig.\ref{fig2}.
It is corresponding to the zero of the  denominator of the equation (\ref{S-Heat1-1}), which yields to the following radius,
\begin{equation}\label{r+c1}
r_{+c}=\frac{2Q}{\sqrt{\Lambda}}.
\end{equation}
Thus, a phase transition occurs in dS space-time, and it does not occur in the AdS space-time.\\
Also, we find that instabilities in dS space-time are depend on value of the black hole charge inspired by conditions (\ref{condition11}) and (\ref{condition22}). We find larger values of the black hole charge yields to the instable phase. These are illustrated by the plots of the Fig. \ref{fig2-1}.\\

Now, by using the relation between negative cosmological constant $\Lambda$ and thermodynamics pressure $P$ as,
\begin{equation}\label{pressure}
P=-\frac{\Lambda}{8\pi},
\end{equation}
we can investigate critical behavior and Van der Waals like phase transition \cite{V1,V2}.\\
Combining condition (\ref{r+c1}), with the pressure (\ref{pressure}) we obtain,
\begin{equation}\label{pressure1}
P_{M}=-\frac{Q^{2}}{2\pi r_{+}^{2}}.
\end{equation}
Now using condition (\ref{horizon1}) to have validity of the first law of thermodynamics in presence of logarithmic correction, we obtain,
\begin{equation}\label{pressure1-1}
P_{M}={\frac {{m}^{4}c^{2} c_1^{2}}{2048\,\pi \,{Q}^{2}}}.
\end{equation}
It is clear that there is no extremum, hence similar to the case of $\alpha=0$ there is no Van der Waals like behavior.
\\
In this paper, we have studied the effects of quantum fluctuations on a BTZ black hole in a massive theory of gravity.
We analyzed the corrections to the thermodynamics of this black hole at a sufficiently small scale.   It had been argued that as
this black hole becomes sufficiently small, we could not neglect the quantum fluctuations. These quantum fluctuations could be expressed
using thermal fluctuations.   We analyzed such thermal fluctuations as perturbations around the equilibrium temperature. We also studied
the stability for such black holes. It was observed that thermal fluctuations can modify the stability of such black holes. Furthermore,
we also analyzed the phase transition for such black holes. We have demonstrated  that the behavior of the BTZ black hole
after taking the thermal fluctuations into consideration is the same as the behavior of the
  BTZ black hole \cite{1509.05481}.
So, we have analyzed the critical point and phase transition
for AdS BTZ black hole in  massive gravity. We also, considered the extremal case of dS space-time.
It would be interesting to analyze such fluctuations for other solutions to such a massive theory of gravity.
It is possible to obtain higher order corrections in the perturbation series, and it would be interesting to analyze the effects
of such corrections on the stability of such black hole solutions. It is also interesting to consider logarithmic correction on the
holographic heat engines \cite{H1, H2}.

\end{document}